\documentclass[10pt]{article}
\usepackage{bbm}
\usepackage[final]{graphics}
\usepackage{amsmath}
\usepackage{amsfonts,amsbsy}
\usepackage{amssymb}
\usepackage{epsfig}
\usepackage{graphicx}
\usepackage{appendix}
\usepackage{slashed}
\usepackage{indentfirst}

\makeatletter \@addtoreset{equation}{section} \makeatother

\def\empile#1\over#2{\mathrel{\mathop{\kern 0pt#1}\limits_{#2}}}

\def\beq{\begin{equation}}
\def\eeq{\end{equation}}
\def\bea{\begin{eqnarray}}
\def\eea{\end{eqnarray}}

\newcommand{\Lb}{\left(}
\newcommand{\Rb}{\right)}
\def\p{{\boldsymbol p}}

\def\d3p{\frac{d^3\p}{(2\pi)^3}E_\p}

\catcode`\@=11

\newcount\@tempcntc
\def\@citex[#1]#2{\if@filesw\immediate\write\@auxout{\string\citation{#2}}\fi
  \@tempcnta\z@\@tempcntb\m@ne\def\@citea{}\@cite{%
        \@for\@citeb:=#2\do%
    {\@ifundefined{b@\@citeb}%
        {\@citeo\@tempcntb\m@ne\@citea%
                \def\@citea{,\penalty\@m\ }{\bf ?}\@warning%
                {Citation `\@citeb' on page \thepage \space undefined}}%
        {\setbox\z@\hbox{\global\@tempcntc0\csname b@\@citeb\endcsname\relax}%%
     \ifnum\@tempcntc=\z@ \@citeo\@tempcntb\m@ne%
       \@citea\def\@citea{,\penalty\@m}%
       \hbox{\csname b@\@citeb\endcsname}%
     \else%
      \advance\@tempcntb\@ne%
      \ifnum\@tempcntb=\@tempcntc%
      \else\advance\@tempcntb\m@ne\@citeo%
      \@tempcnta\@tempcntc\@tempcntb\@tempcntc\fi\fi}}\@citeo}{#1}}%

\def\@citeo{\ifnum\@tempcnta>\@tempcntb\else\@citea
  \def\@citea{,\penalty\@m}%
  \ifnum\@tempcnta=\@tempcntb\the\@tempcnta\else
   {\advance\@tempcnta\@ne\ifnum\@tempcnta=\@tempcntb \else
\def\@citea{--}\fi
    \advance\@tempcnta\m@ne\the\@tempcnta\@citea\the\@tempcntb}\fi\fi}

\catcode`\@=12

\begin{document}

\title{\bf Production of Photons and Dileptons in the Glasma}
\author{Mickey Chiu${}^{(1)}$,  Thomas K. Hemmick${}^{(2)}$,  Vladimir Khachatryan${}^{(2)}$, Andrey Leonidov${}^{(3)}$\footnote{Also at ITEP, Moscow and MIPT, Moscow.},\\
 Jinfeng Liao${}^{(4,5)}$, Larry McLerran${}^{(1,5)}$}

\maketitle

\begin{enumerate}

\item Physics Department, Bldg. 510A, Brookhaven National Laboratory,
   Upton, NY 11973, USA
  \item  Physics Department, Stony Brook University, Stony Brook, NY 11794-3800, USA\
\item Lebedev Physical Institute, Leninsky Pr. 53, 119991 Moscow, Russia
\item   Physics Department and Center for Exploration of Energy and Matter,
Indiana University, 2401 N Milo B. Sampson Lane, Bloomington, IN 47408, USA
 \item RIKEN BNL Research Center, Bldg. 510A, Brookhaven National Laboratory,
   Upton, NY 11973, USA
\end{enumerate}

\begin{abstract}
 We study the production of photons and dileptons during the pre-equilibrium Glasma stage in heavy ion collisions and discuss the implications in light of the  PHENIX data.
 We find that the measured distributions of such electromagnetic emissions, while having some features not well understood if hypothesized to entirely arise from a thermalized Quark-Gluon Plasma, have some qualitative  features that might  be described after including effects from a thermalizing Glasma. The shape and centrality dependence of the transverse momentum spectra  of the
 so-called "thermal photons" are well described. The mass and transverse momentum dependence of intermediate mass dileptons also agree with our estimates. 
 The low transverse momenta from which the excessive  dileptons (in low to intermediate mass region) arise is suggestive of emissions from a Bose condensate.
 We also predict the centrality dependence of dilepton production. Uncertainties in the current approach and improvements in the future are discussed. 
% Our simple estimates are not sufficiently precise to determine the overall normalization
% of either the dilepton or photon production, nor, as is the case for a thermalized Quark-Gluon Plasma, to explain the observed flow of the photons. 
 \end{abstract}

%\preprint{ }

\section{Introduction }

Two  traditional probes of matter produced in heavy ion collisions
are photons and dileptons\cite{Shuryak:1978ij}-\cite{Feinberg:1976ua}.  Photons and dileptons, while  produced since very early times in the collisions, propagate through the produced
matter  largely without interaction due to the small electromagnetic cross sections. Therefore they provide unique access to the history of the quark gluon matter from early times, including  the pre-equilibrium evolution. 

Measurements of such photons and dileptons produced in collisions at RHIC were first reported by the PHENIX experimental collaboration with quite surprising results \cite{:2008fqa}-\cite{Adare:2009qk}.  There is a large excess of photons in the transverse momentum range of $1-3~GeV$ for central gold-gold collisions.  This excess far exceeds that due to direct photons and has been interpreted by the PHENIX collaboration to represent photons produced by a Quark-Gluon Plasma with a temperature in excess of the deconfinement temperature.
Recently PHENIX has also measured the flow (or more precisely the azimuthal anisotropy) of such photons  and found it to be sizable and exceeding the expectation from the hydrodynamic expansion of a thermalized Quark-Gluon  Plasma \cite{Adare:2011zr}.  Furthermore an excess of dileptons in the mass range $ 100~ MeV \le M \le 1~GeV $ is also present in the PHENIX data.  Hydrodynamic computations assuming a thermalized Quark-Gluon plasma fail to reproduce the magnitude of this dilepton excess.  In addition, the slope of the $k_T$ distribution for such excessive dileptons in the relevant invariant mass bins is in the range of few hundred MeV,  much less than the corresponding mass of the dileptons: this fact is in sharp contrast with the typical kinematics $k_T \sim M$ if  such pairs were to arise either from a thermal emission from a Quark Gluon Plasma or from semi-hard processes.
 
More recently the STAR experimental collaboration has searched for an excess of low to intermediate mass dileptons and the reported results are not
in apparent accord with those of PHENIX \cite{wang}-\cite{zhao}.  The excess from STAR data is less than that seen in PHENIX and might be 
accommodated by a themalized Quark-Gluon Plasma hypothesis.  At the moment it is not clear yet whether the STAR measured enhancement  of these pairs  have their origin in low transverse momentum region.   The ALICE collaboration has reported seeing an excess of photons for Pb-Pb collisions 
at 2.76 TeV in the same transverse momentum range as that of PHENIX, but have not yet reported results concerning flow\cite{safrik}.

Whether the ultimate origin of the  ``thermal photons" and dilepton excess may be entirely attributed to  a thermally equilibrated Quark-Gluon Plasma with high temperature will remain to be seen, and a necessary first step would be the experimental clarification of the current tension between the PHENIX and STAR data. The purpose of the present study is to explore an alternative hypothesis based on possible important contribution to the electromagnetic emissions in the pre-equilibrium matter, the Glasma. As shown in a series of recent studies~\cite{Kovner:1995ja}-\cite{Lappi:2006fp}, the Glasma begins in the earliest stages of heavy ion collisions as an ensemble of longitudinal color electric and color magnetic lines of flux, and then decays into a far-from-equilibrium gluon-dominant matter. Owing to the high occupancy of gluons that coherently amplifies scattering,  the resulting Glasma appears strongly interacting during the (presumably long) time toward thermalization even though the coupling is weak~\cite{Dusling:2010rm}-\cite{Kurkela:2012hp}. It may even develop a transient yet robust Bose-Einstein Condensate of gluons as recently suggested in \cite{Blaizot:2011xf} with supportive evidences from classical statistical lattice simulations~\cite
{Epelbaum:2011pc,Berges:2011sb,Berges:2012us,Berges:2012ev}.  
The name Glasma implies its nature as the matter in between that of the Color Glass Condensate
~\cite{McLerran:1993ni} and that of the thermally equilibrated Quark-Gluon Plasma, which can be thought of as a strongly interacting but unequilibrated form of  Quark-Gluon Plasma.

In this paper, we will take a first step to make rough estimates of the electromagnetic emission properties of this Glasma.   While our estimates are not as detailed as those that exist for the thermalized Quark-Gluon Plasma from years of efforts, we will find nevertheless, that the Glasma appears to have  the correct qualitative and semi-quantitative features to explain some interesting aspects of the  PHENIX photons and dileptons data that would otherwise be difficult to understand.  In this paper a number of approximations will be made due to our currently incomplete knowledge of the Glasma.  Our estimates will mostly focus on the dependence of various distributions upon centrality, transverse momentum and mass.   The overall normalization of these effects is not determined within our current crude approximations but rather fixed from data.  This may not be surprising by recalling that determining normalization of rates was very difficult for the thermalized Quark-Gluon Plasma and involved much effort over many years.  A further weakness of our present analysis is that we will treat only 1+1 dimensional
 expansion of the Glasma and as such cannot estimate transverse flow effects.  In addition,  we are aware of the situation that the existence of a transient Bose condensate
 in the Glasma is still under intensive ongoing investigations and so far neither fully established nor ruled out as consequence of the complicated dynamics in the Glasma, and thus the conclusions concerning dileptons may be correspondingly weakened. Finally one should be cautious about the current dispute on data for dileptons between STAR and PHENIX.
 
 Given all these uncertainties, what do we learn from this analysis?  
 
 For the photons:
 
 \begin{itemize}
 
 \item{The Glasma hypothesis yields a simple and robust estimate for the dependence of photon production on centrality as a consequence
 of geometric scaling of the emission amplitudes. As will be shown, the PHENIX photon data very well satisfies such geometric scaling.  }
 
 \item{The Glasma hypothesis can generate the correct shape of the transverse momentum spectrum.
 %  It is somewhat more flexible than a themalized QGP hypothesis.
 }
 
 \item{The photon flow, while not computable in our current approach, may  arise naturally from the pre-equilibrium flow patterns that can be generated in Glasma, earlier
 than is usually assumed for the initial conditions for the thermalized QGP.  In the Glasma the  quarks become substantial only till relatively late times of the evolution, thus allowing flow to establish. } 
 
 \end{itemize}

 For the dileptons:
 
 \begin{itemize}
 
 \item{The transverse momentum and mass spectra  can be described within the Glasma hypothesis.}
 
\item{ The observation by PHENIX that the low to intermediate mass enhancement arises from anomalously small transverse
 momentum region has a natural interpretation as emissions from a condensate which  in a fluid at rest produces particles with zero transverse momentum.}
 
\item{ Geometric scaling is predicted for the centrality dependence of the dilepton spectra. }

\end{itemize}

The rest of this paper is organized as follows:  in the second section we will review those recently developed results for the Glasma which will be needed to estimate the photon and dilepton rates;  in the third section, we will then use these results to estimate the photon and dilepton rates from Glasma emissions, accounting for the time evolution of the Glasma; in the fourth section, we compare our results with the data from the PHENIX collaboration; finally in the last section,
we will draw our conclusions and end by discussing a number of caveats in our present estimates.

 \section{Review of Relevant Properties of the Thermalizing Glasma}

In this section, we briefly review a recently proposed scenario for the thermalization process in the Glasma. The results relevant to our  discussion of electromagnetic production will be presented, while all the details can be readily found in \cite{Blaizot:2011xf}.

  During the earliest stage of the evolution of the Glasma, $0 \le t \sim 1/Q_{sat}$ where $Q_{sat}$ is the gluon saturation momentum, the gluonic degrees of freedom are largely coherent longitudinal color electric and color magnetic fields.
Gluons, in the sense of particles,  are being produced  from the classical evolution of color electric and color magnetic fields, and these gluons produce a distribution that is approximately isotropic in momentum space, due to plasma instabilities.  Not many quarks are present since they are produced by quantum fluctuations in the gluon field, and this is suppressed by a power of $\alpha_s$.   The QCD coupling constant is small if the gluon saturation momentum is large compared to the QCD scale, which we shall assume.

We shall not discuss the photons and dileptons produced at this earliest time.  Instead, we will concentrate on the time interval $1/Q_{sat} << t << t_{therm}$ where $t_{them}$ is the thermalization time.  During this time interval, the quark density increases to a value of the order of the gluon density, and is no longer suppressed.  Since electromagnetic particle production ultimately arises from the electromagnetic charges of quarks,  it is plausible that
the production for $t \sim 1/Q_{sat}$ might not be important.  Further, we will concentrate on transverse momentum and mass scales where we expect that the effects of the evolution to a thermalized distribution are enhanced.

We assume the gluon distribution function is of the form
\begin{equation}
  f_g = {\Lambda_s \over {\alpha_s p}} F_g(p/\Lambda)
\end{equation}
In this equation, $p$ is the gluon momentum.  $\Lambda_s$ is the momentum scale at which the gluons are maximally coherent and is time dependent.  At the earliest times $\Lambda_s(t_0) \sim Q_{sat}$.
$\Lambda$  is a time dependent ultraviolet cutoff, which at the earliest time coincides with $\Lambda_s$ i.e. $\Lambda(t_0) = \Lambda_s(t_0)$.  The scale $\Lambda$ however continuously separates from $\Lambda_s$ during the course of thermalization, and upon equilibration becomes the initial  temperature for the Quark-Gluon Plasma $\Lambda(t_{therm}) \sim T_i$. The soft scale $\Lambda_s$, on the other hand, becomes the non-perturbative ``magnetic scale''  \cite{Liao:2006ry} in the thermalized plasma $\Lambda_s(t_{therm}) \sim \alpha_s\, T_i$. The thermalization is therefore accomplished by splitting apart these initially overlapping  momentum scales by $\alpha_s$ parametrically, and the corresponding time is determined by the following requirement
\begin{equation}
    \Lambda_s(t_{therm}) \sim \alpha_s \Lambda(t_{therm})
\end{equation}
To achieve such separation takes parametrically long time in the very high energy limit and could take considerable time even at the RHIC energy.

A very important consequence of the saturation is that the phase space for the gluons is initially over-occupied
\begin{equation}
  n_g/\epsilon_g^{3/4} \sim 1/\alpha_s^{1/4}
\end{equation}
where $n_g$ is the number density of gluons, and $\epsilon_g$ is the energy density in the gluons.  For a thermally equilibrated Bose system, it is necessary that this ratio be less than a number of the order 1.
  It is therefore plausible that in addition to the gluons, a Bose-Einstein condensate could be developed with time by ``absorbing'' the large number of excessive gluons into zero momentum state (provided that the inelastic processes are not fast enough to sufficiently reduce the number of gluons prior to thermalization). Such a condensate would be of the form
\begin{equation}
    f_{cond} = n_{cond} \delta^{3} (p)
\end{equation}
The condensate can be thought of as many gluons compressed into a color singlet and spin singlet configuration that are highly coherent and have zero momentum.  We expect that the effective masses of the gluons in the condensate should be of the order of the Debye scale, which is
\begin{equation}
   M^2_{Debye} \sim \Lambda \Lambda_s
\end{equation}
This is because when the condensate decays it must produce real time excitations and the minimum
mass scale for such excitations is the Debye mass.  The Debye mass will also act as an infrared cutoff in various dynamical processes.

It was shown in \cite{Blaizot:2011xf} that the time evolution is dominated by the gluon density, and that there may be some fixed
asymmetry between the typical transverse and longitudinal momentum scales characterized by a parameter
$\delta$.  The parameter $\delta$ is defined in terms of the longitudinal pressure
\begin{equation} \label{eqn_PL}
  P_L = \delta \, \epsilon
\end{equation}
where $0 \le \delta \le 1/3$, with $\delta=0$ and $\delta=1/3$ corresponding to the free-streaming (thus maximal anisotropy between the longitudinal and transverse pressure) and the isotropic expansion, respectively.
The time evolution of the scales $\Lambda_s$ and $\Lambda$ were found to be
\begin{equation}
     \Lambda_s \sim Q_s \left( {t_0 \over t} \right)^{(4+\delta)/7}
\end{equation}
and
\begin{equation}
     \Lambda \sim Q_s \left( {t_0 \over t} \right)^{(1+2\delta)/7}
\end{equation}
This can be translated into the gluon density and the Debye mass as
\begin{equation}
  n_g \sim {Q_{sat}^3 \over \alpha_s} \left( {t_0 \over t} \right)^{(6+5\delta)/7}
\end{equation}
and
\begin{equation}
  M^2_{Debye}  \sim Q_{sat}^2  \left( {t_0 \over t} \right)^{(5+3\delta)/7}
\end{equation}
The thermalization time is given by
\begin{equation}
  t_{therm} \sim t_0 \left( 1 \over \alpha_s \right)^{7/(3-\delta)}
\end{equation}

It is difficult to determine the time evolution of the gluon condensate density without fully addressing the inelastic processes. Nevertheless one may assume an approximation transport equation of the form
\begin{equation}
  {d \over {dt}} n_{cond} = - {a \over t_{scat}} n_{cond} + {b \over t_{scat}} n_{g}
\end{equation}
Here $t_{scat} \sim t$ is the universal scattering time for the Glasma and  $a$ and $b$ are constants of order 1.  The first term represents the decay of condensate due to inelastic processes, while the second term reflects the ``feeding'' into condensate from the over-occupied gluons.  (This equation is for illustration shown for a non-expanding medium, but analogous results are easy to derive for an expanding medium.)  Under the above approximation it can be deduced  that either the condensate decreases more slowly than the gluon density if $a < b$, or the condensate density stays at the order of the gluon density if $a\ge b$. In either case one has $n_{cond} \ge n_g$, and for simplicity we will use the following assumption for later discussions:
\begin{equation}
  n_{cond} = \kappa\, n_{gluon}
\end{equation}
where $\kappa$ is a constant of order 1.

Finally, to complete our description, we need the quark number density that is
\begin{equation}
    f_q =  F_q(p/\Lambda)
\end{equation}
so that the quark number density is
\begin{equation} \label{eqn_ng_nc}
   n_q \sim \Lambda^3
\end{equation}
At the earliest times $n_q \sim \alpha_s n_g << n_g$, but at late time the two densities approach each other, i.e. $n_q \sim n_g$.

\section{Electromagnetic Particle Production from the Glasma}

Now let us estimate the rate of photon production from the Glasma.  Recall that the quark number density,
up to an overall constant is identical to that for the quark number density in a Quark Gluon Plasma
with the replacement $ \Lambda \rightarrow T$.  The computations of the rate for photon production at finite temperature
are reviewed in Ref. \cite{Srivastava:2001hz}.   For thermal emission from a Quark Gluon Plasma in a fixed box, the result is
\begin{equation}
   {{dN} \over {d^4x dy d^2k_T}} = {{\alpha \alpha_s} \over {2\pi^2}} T^2 e^{-E/T} h(E/T)
\end{equation}
where $h$ is a slowly varying function of $E/T$ of the order one.  The factor of $\alpha_s$ arises
from the interaction of quarks with the medium in the photon production process.  (This formula and the ones
that follow are evaluated in the local rest frame of the fluid, and require generalization for use
in a boosted frame.)

In  the Glasma, this is compensated for  by the high gluon density $\sim 1/\alpha_s$ associated with the coherence of the Glasma.  For Glasma emission, we shall use a simplified form of this equation,
\begin{equation}
  {{dN} \over {d^4x dy d^2k_T}} = {\alpha \over \pi} \Lambda_s \Lambda  g(E/\Lambda) \label{photratefb}
\end{equation}
Here, $g$ is a function of order one that cuts off when then energy of the photon is of the order
of the UV cutoff scale $\Lambda$.  This form follows from dimensional reasoning, and the fact that the overall rate must be proportional to the electromagnetic coupling. The factor of $\Lambda$ is analogous to the
temperature factor for thermal emissions.  There is a factor of $\Lambda_s/\alpha_s \Lambda$ relative to the naive generalized thermal formula.  This factor arises because one of the external legs of the diagram that induces photon emissions  couples to
a coherent Glasma gluon and this has a distribution function proportional to $\Lambda_s/\alpha_s$.
The factor of $g(E/\Lambda)$ occurs because $\Lambda$ is the largest momentum scale in the problem and quarks always have a typical momentum scale of order $\Lambda$.  Whether the gluon arises from thermal gluons or from the gluon condensate is not important, since we will assume the density of gluons and the density of the condensate are the same. This will only affect the external line factors but not the dependence of $g(E/\Lambda)$, because it is the largest momentum scale, i.e. that of the quarks, that determines such dependence. Note that at thermalization when $T \sim \Lambda \sim \Lambda_s/\alpha_s$, this formula reduces to that for thermal emission. A heurisic derivation of Eq. (\ref{photratefb}) is given in the paragraph \ref{apphot} in the Appendix.

%{
%\begin{eqnarray}
%\frac{\Lambda_s}{Q_s} \sim \left(\frac{\Lambda}{Q_s}\right)^{\frac{4+\delta}{1+2\delta}}
%\end{eqnarray}
%}

To obtain the overall rate, we need to integrate over longitudinal coordinates.  We assume that the early time expansion is purely longitudinal, and that in the integration the space-time rapidity is strongly correlated with that of the momentum space-rapidity.  We then have that
\begin{equation}
{{dN} \over {d^2r_T dyd^2k_T}} \sim \alpha \int tdt \Lambda_s \Lambda g(k_T/\Lambda)  \label{rate}
\end{equation}

Using the result of the previous section for the time dependence of the scales $\Lambda$ and $\Lambda_s$, we have
\begin{equation}
   t dt = \kappa^{\prime}~  {{d\Lambda} \over \Lambda}~ {1 \over Q_{sat}^2} \left(Q_{sat} \over \Lambda
   \right)^{14/(1+2\delta)}
\end{equation}
The constant $\kappa^{\prime}$ is of order 1.\\

Doing the integration over $\Lambda$ in Eqn.(\ref{rate}),
we find that
\begin{equation}
   {{dN} \over {d^2r_T dy d^2k_T}} \sim \alpha \left(Q_{sat} \over k_T \right)^{\frac{9-3\delta}{1+2\delta}}
\end{equation}

Now integrating over $d^2r_T$,  and identifying the overlap cross section as proportional to the number of participants, we finally obtain
\begin{equation}
       {{dN_{\gamma}} \over {dy d^2k_T}} =  ~\alpha ~R_0^2 ~N_{part}^{2/3}   \left(Q_{sat} \over k_T \right)^\eta
\end{equation}
where $\eta=(9-3\delta)/(1+2\delta)$.
The factor of $N_{part}^{2/3}$ arises because the number of participants in a collision proportional
to the nuclear volume $R^3 \sim N_{part}$
Here $R_0$ is a constant with dimensions of a length.  It should be of order $1~fm$, but cannot be determined precisely due to the crude approximations made.   The power of $Q_{sat}/k_T$ ranges from
\begin{equation}
      9 \ge \eta \ge 24/5
\end{equation}
with the  two limits $\eta=9$ and $\eta=24/5$ corresponding to $\delta=0$(maximal anisotropy) and $\delta=1/3$ (isotropic expansion) respectively.  Note that in this formula, once the power of
$k_T$ has been determined from experiment, then using that $Q_{sat}^2 \sim N_{part}^{1/3}$, we have
that the cross section scales as $N_{part}^{2/3 +\eta/6}$.  This is a very rapid dependence on the number of participants.

It is important to note that in the derivation of this result, we have assumed that the largest part of
the contribution when integrating over $\Lambda$ does not come from the end points of the integration.
If the end points become important, then the physics either from the earliest times (hard processes) or
from the thermalized Quark Gluon Plasma will become important.  For the case of a Quark Gluon Plasma, the dominant region of integration is $k_T \sim 6 T$.  The smallest possible value for  $\Lambda$ would be of the order of the QCD transition temperature, and at RHIC energies the highest possible value for $\Lambda$ shall be around $1\rm GeV$.  These considerations are therefore valid at best for photon production in the range of $1~\rm GeV \le k_T \le 10~GeV$.

The analysis of dilepton production is more complicated  because there are two sources of dileptons.  The first is due to annihilation of quarks in the Glasma.  For this contribution, the rate can be determined by dimensional reasoning to be\footnote{For a heurisic derivation of Eq. (\ref{dilratefb}) see paragraph \ref{apdil} in the Appendix.}
\begin{equation}
  {{dN_{DY}} \over {d^4x dM^2}} = \alpha^2 \Lambda^2 g^{\prime}(M/\Lambda) \label{dilratefb}
\end{equation}
The evaluation of this contribution follows as above and gives
\begin{equation} \label{eqn_pair}
{{dN_{DY}} \over {dy dM^2}} \sim \alpha^2 R_0^{\prime 2} ~N_{part}^{2/3} \left(Q_{sat} \over M \right)^\eta
\end{equation}
 with $\eta=4(3-\delta)/(1+2\delta)$ taking values $12\ge \eta \ge 32/5$ for $0\le \delta \le 1/3$.
Unfortunately, the derivation of this result requires very massive dileptons.  Analogous to the limits on $k_T$
for direct photon production, such contribution would be in the mass range greater than $1~\rm GeV$ in which a variety of other processes such as charm particle decays would obscure a Drell-Yan signal. Also note that the most interesting ``excess'' seen in the RHIC data appears in the range between a few hundred $\rm MeV$ and about $1~\rm GeV$.

\begin{figure}[tb]
\vspace{0.0cm}
\begin{center}
\scalebox{1.0}[1.0] {
\hspace{0.2cm}
  \includegraphics[scale=.55]{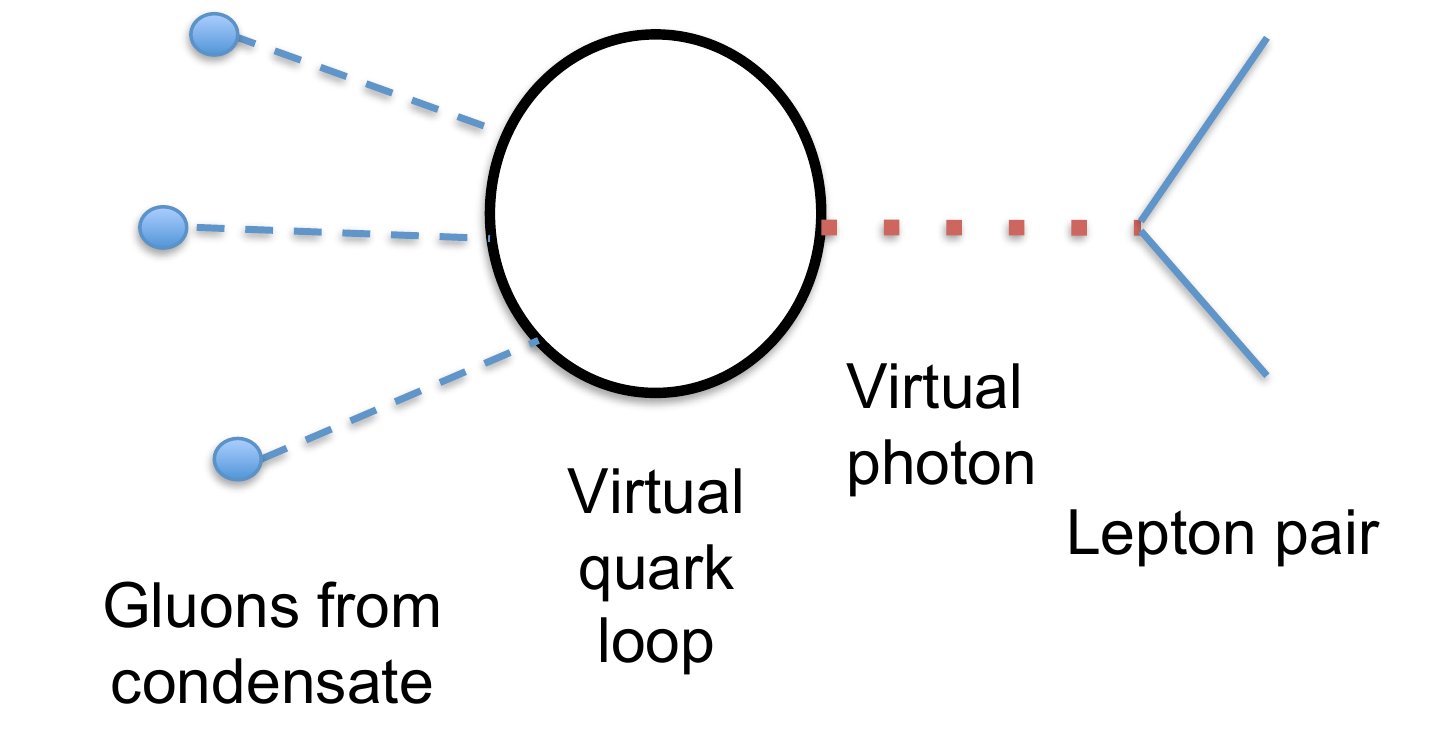} }
\end{center}
\caption{Three gluons from the condensate annihilate into a virtual quark loop, that subsequently decays into a virtual photon and then into a dilepton.
}
\label{fig:threegluoni}
\vspace{0.2cm}
\end{figure}

Let us however consider another process.  This process is the annihilation of gluons into a quark loop from which the quarks then subsequently decay into a virtual photon and eventually the dilepton: see the illustration in the Fig.\ref{fig:threegluoni}.  Such a virtual process is naively suppressed by factors
of $\alpha_s$.  Here however, the gluons arise from a highly coherent condensate, and the corresponding
factors of $\alpha_s$ are compensated by inverse factors $1/\alpha_s$ from the coherence of the condensate.   In other words, the usual power counting for diagrams in terms of $\alpha_s$ has to be changed when the coherent condensate with high occupation is present.

This annihilation process from a condensate has a distinctive feature.  The condensate gluons have nearly zero total momentum in a co-moving frame.  All of the gluon momentum is acquired by collective flow, and hence
the produced dileptons will have a small transverse momentum  which could be much smaller compared with the pair mass.  In contrast, for usual thermal production processes
 as well as hard particle production processes, the typical transverse momentum of the produced pair is of
the order of the dilepton mass.

Here we estimate the rate for the three-gluon decay of the condensate into a dilepton.  On dimensional grounds, we expect that
\begin{eqnarray}
 {{dN_{C\to DY}} \over {d^4x dydM^2}} = \alpha^2 {{(\alpha_s n_{gluon})^3} \over M_{Debye}^7}
 g^{\prime\prime}(M/M_{Debye})
\end{eqnarray}
We are assuming the condensate density is of the order of the gluon number density as in Eq.(\ref{eqn_ng_nc}).  We are also assuming
that the typical scale for the energy of gluons in the condensate is of order the Debye mass.  It is also implicitly
assumed that the condensate is unstable with respect to decay but gets  stabilized by processes that
build up the condensate due to the over-occupation of gluonic states.
Note that in the scaling Glasma we have $M_{Debye} \sim \sqrt{\Lambda\Lambda_s}\sim Q_s (t_0/t)^{(5+3\delta)/14}$.
Note that  due to the external gluon lines,
 there is a correction to the result in Eq.(\ref{eqn_pair})  on the order of
  $(\alpha_s n_{gluon})^3/M_{debye}^9$. The coupling $\alpha_s$ from each leg will be canceled by the high density $1/\alpha_s$ from each leg and at the end one obtains an overall multiplicative factor $(\alpha_s n_{gluon}/M_{debye}^3)^3 \sim (t/t_0)^{3(3-\delta)/14}$. When the integration over time is concerted into that of factors of mass, we get the following result
  \begin{equation} \label{eqn_dilep_int}
  {{dN_{C\to DY}} \over {dydM^2}} \sim \alpha^2 R_0^{\prime 2} N_{part}^{2/3} \left( {Q_{sat} \over M} \right)^{\eta^{\prime}}
\end{equation}
where the exponent now becomes $\eta'$ given by
 \begin{equation}
        \eta^\prime_{perturbative} = \frac{9(3-\delta)}{5+3\delta}
 \end{equation}
It is valued in the range $27/5 \ge \eta^\prime_{perturbative} \ge 4$ corresponding to $\delta = 0$ for maximal anisotropy and $\delta=1/3$ for isotropic expansion.

 Because the time evolution of the Debye mass is much
 more rapid than is that for the ultra-violet cutoff scale $\Lambda$, we expect that the range where this formula
 applies is in a range of masses significantly smaller than is the case of the photons,
 and Drell-Yan emission for quarks.  It is not unreasonable to expect that this formula works in the range somewhat below $1 ~GeV$ but may cut off at some small mass of order few hundred $MeV$.  A detailed computation of the rate and determination of behavior near such a cutoff would be useful, since the cutoff is a measure of the temperature at which the condensate disappears, which is presumably the thermalization temperature.

The results we present above for three-gluon annihilation into a quark loop can and should be corrected for
multiple gluon annihilation.  For such soft gluon attachments, we believe these effects can, with some work, be analytically summed.  Since more gluon attachments will increase the powers of time-dependence on the external legs, this will lead to a steepening of the dependence on the mass scale.  The generic feature of geometrical scaling, that the distribution is a function of the form
\begin{equation}
   {{dN_{C\to DY}} \over {dydM^2}} \sim \alpha^2 R_0^{\prime 2} N_{part}^{2/3} F_{DY}(Q_{sat}/M)
 \end{equation}
 will not be modified, as this follows entirely from dimensional reasoning.  The distribution from the condensate will also come from small momentum.  We see that generic features of the distribution that we wish to extract will remain, although the shape of the curve in $M$ would have significant modification.
 As a practical matter, one needs to do the integration over the space-time history much more accurately
 for the mass range seen at RHIC energies, since this range extends to rather low mass values.

  We end by discussing a few theoretical issues in the above estimation. With the crude approximation above, we have estimated the contribution from one flavor of massless quark.  The amplitude for this contribution is proportional to the quark charge.  If we have multiple massless flavors, we have
 it proportional to $\sum_i e_i$.  For the three light flavors $u,d,s$, this sum   would vanish provided they are strictly degenerate in mass.  A non-vanishing contribution would survive when the mass for the strange quark becomes relevant (e.g. when the scales are low), and/or a contribution from the charm quark becomes relevant (e.g. when the scales are high).
 Which contribution is dominant depends upon the scale of the dilepton mass.  These effects will generate an overall suppression and may introduce a non-trivial shape into the
mass distribution.  For example, the charm quark contribution could make the distribution harder i.e. less steep. 

Another concern is related to the spin states of the gluons in the condensate. The diagram in Fig.\ref{fig:threegluoni} computes essentially the EM vector current correlator, and since the condensate gluons have zero spatial momentum the spatial structure will have to arise from the gluon spin indices. A nonzero contribution from the diagram is obtained if the gluon spin states are either incoherent and trivially averaged as is the case in thermal QGP (which we think shall be the case) or are coherently in a uniform spin orientation (thus breaking spatial isotropy).  These issues would require further works for clarification.

 \section{Phenomenology of Photon and Dilepton Emission}

 In this section we will present phenomenological formulae for electromagnetic emissions that follow from the theoretical results described in the previous sections, and compare the results with the experimental data on photons and dileptons from the PHENIX collaboration.

 \subsection{Photons}

 We begin with Eq.\,$(21)$ for the photon yield
\begin{equation}
{dN_{\gamma} \over dy\,d^{2}k_{T}} \sim \alpha R_{0}^{2}N_{part}^{2/3}\Lb {Q_{sat} \over k_{T}} \Rb^{\eta}\,,
\label{eqn_phot1}
\end{equation}
where \,$\eta = (9 - 3\delta)/(1 + 2\delta)$.\, The power of \,$Q_{sat}/k_{T}$\, ranges from
\,$9 \geq \eta \geq 24/5$\, (corresponding to \,$0 < \delta < 1/3$,\, respectively). We use that
\begin{equation}
Q_{sat}^{2}(k_{T}/\sqrt{s}) =  Q_{0}^{2}\Lb \sqrt{s}\times 10^{-3} \over {k_{T}} \Rb^{\lambda}\,,
\label{eqn_mom1}
\end{equation}
where the $\lambda$ is a parameter characterizing the growth of the saturation momentum with decreasing $x$. These lead to a result paralleling that of the analysis for $pp$ scattering in Ref.\cite{McPras1}
\begin{equation}
{dN_{\gamma} \over dy\,d^{2}k_{T}} \sim \alpha R_{0}^{2}N_{part}^{2/3}\Lb {\sqrt{Q_{0}^{2}\Lb
\sqrt{s}\times 10^{-3}
/ {k_{T}} \Rb^{\lambda}} \over k_{T}} \Rb^{\eta} \sim
\alpha R_{0}^{2}N_{part}^{2/3} {\Lb Q_{0}^{2}\Lb \sqrt{s}\times 10^{-3} \Rb^{\lambda} \Rb^{\eta/2}
\over k_{T}^{\,\eta\,(1+\lambda/2) }}\,.
\label{eqn_phot2}
\end{equation}
Phenomenologically, one would expect a range $0.2 \le \lambda \le 0.35$. In this paper we will examine this  range of parameters and find the best fitting results.

Based on the above, we will use the following phenomenological formulae for parameterizing the contribution from Glasma evolution to the photon production:
\begin{equation}
F(\lambda,\eta) \equiv C_\gamma\, N_{part}^{2/3} \times {\left [ Q_{0}^{2}\Lb \sqrt{s}\times 10^{-3} \Rb^{\lambda} \right ]^{\eta/2}
\over k_{T}^{\,\eta\,(1+\lambda/2) }}\,.
\label{eqn_F}
\end{equation}
where the constant coefficient $C_\gamma \propto \alpha R_0^2$ can be determined by fitting at one centrality bin and then applied to all other centralities.
For comparison with data, one also needs to include the photon production from the initial $pp$ collisions (without any medium effect). Such production can be described by properly scaled-up pQCD results for $pp$ collisions. We use the Hagedorn function for parameterizing this ``trivial'' contribution:
 \begin{equation}
 G = T_{AA} \times \frac{A_{pp}}{(1+k_T^2/b)^n}
 \label{eqn_G}
 \end{equation}
 where $T_{AA}$ is the Glauber nuclear overlap function depending on centrality.
This part of the contribution has been studied with the parameters determined to be \,$A_{pp} = 0.0133264\,{\rm mb \; GeV^{-2}}$,\, \,$b = 1.5251\, \rm GeV^{2}$\,
and \,$n = 3.24692$.
The phenomenological formula for total photon production will therefore be a sum of the two contributions $F+G$.

The data we aim to describe will be the invariant yield in $Au$-$Au$ collisions
at \,$\sqrt{S_{NN}} = 200\,GeV$\, of direct photons at centralities \,$0$-$20^{\circ\!\!}/_{\!\!\circ}$,\,
\,$20$-$40^{\circ\!\!}/_{\!\!\circ}$\, and \,$0$-$92.2^{\circ\!\!}/_{\!\!\circ}$ (Min. Bias) as a function
of $k_{T}$ determined  from the PHENIX measurement: see Fig.\,$34$ of Ref.\,\cite{Adare:2009qk}. The strategy is the following: for given values
of $\lambda$ and $\eta$ in Eq.(\ref{eqn_F}), we will fix the coefficient $C_\gamma$ from the $0$-$20^{\circ\!\!}/_{\!\!\circ}$ case and test how well the formula describe the data at the other two centrality choices. This will provide a critical test of the geometric scaling properties of the present model.

\begin{figure}[t]
\vspace{0.0cm}
\begin{center}
\scalebox{1.0}[1.0] {
\hspace{0.2cm}
  \includegraphics[scale=.55]{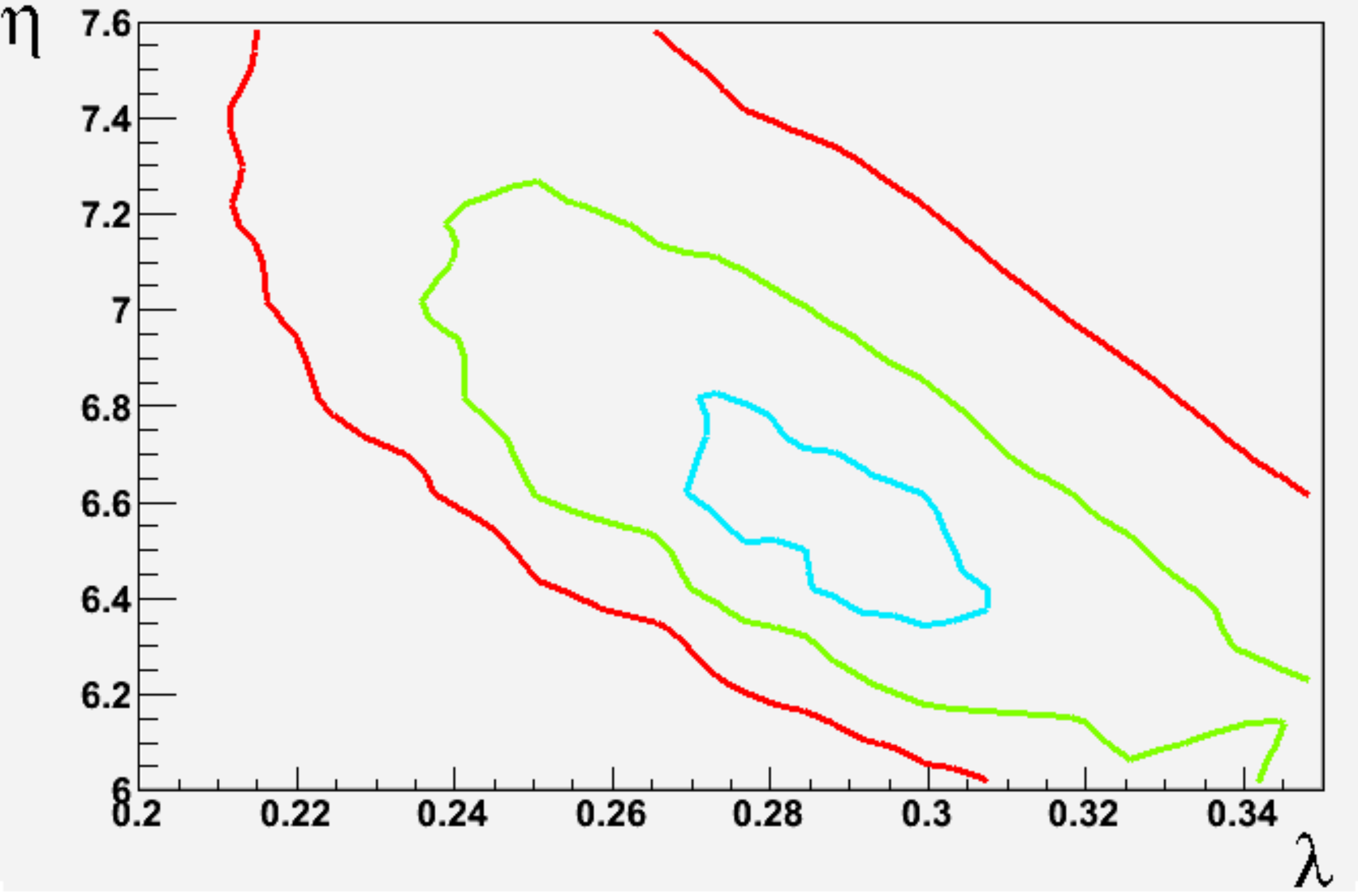} }
\end{center}
\caption{The $\chi^2$/d.o.f. analysis in the $\lambda-\eta$ parameter space for the PHENIX photon data in three centrality bins by fitting with the present model, with the blue, green, and red contours indicating $1$-, $2$-, and $3$-$\sigma$ errors, respectively  (see text for more details). }
\label{photons_chi}
\vspace{0.2cm}
\end{figure}

We now discuss the various parameters involved in the comparison. \\
(1) For the key parameters $\lambda$ and $\eta$ in Eq.(\ref{eqn_F}), we test a wide range of choices for $0.2 \le \lambda \le 0.35$ and $24/5 \le \eta \le 9$. For each specification of $\lambda$ and $\eta$ values, we can do the fitting for photon data at all centralities and evaluate the corresponding $\chi^2$/d.o.f. which will allow us to find the regions of $\lambda$ and $\eta$ for the best fitting results. \\
(2) For $N_{part}$ and $T_{AA}$, we use the Glauber model calculation from PHENIX for these centralities:
$<\!N_{part}\!> = 279.9$ and $T_{AA} = 18.55\,mb^{-1}$\, at \,$0$-$20^{\circ\!\!}/_{\!\!\circ}$;
$<\!N_{part}\!> = 140.4$ and $T_{AA} = 7.065\,mb^{-1}$\, at \,$20$-$40^{\circ\!\!}/_{\!\!\circ}$;
$<\!N_{part}\!> = 109.1$ and $T_{AA} = 6.14\,mb^{-1}$\, at \,$0$-$92.2^{\circ\!\!}/_{\!\!\circ}$ (Min. Bias).\\
(3) For the saturation scale $Q_0$ in Eq.(\ref{eqn_F}), we determine its value at various centralities and beam energies by using the scaling properties $Q_0^2 \propto N_{part}^{1/3}$ and $Q_0^2 \propto (\sqrt{s})^{\lambda/(1+\lambda/2)}$ (see \cite{KLN}\cite{KN} for details), which gives the following values $Q_0^2(0 \mbox{-} 20^{\circ\!\!}/_{\!\!\circ})=1.895\rm GeV^2$, $Q_0^2(20 \mbox{-} 40^{\circ\!\!}/_{\!\!\circ})=1.490\rm GeV^2$, and $Q_0^2(0 \mbox{-} 92.2^{\circ\!\!}/_{\!\!\circ})=1.384\rm GeV^2$ to be used in our case.

\begin{figure}[tb]
\vspace{0.0cm}
\begin{center}
\scalebox{1.0}[1.0] {
\hspace{0.2cm}
  \includegraphics[scale=.88]{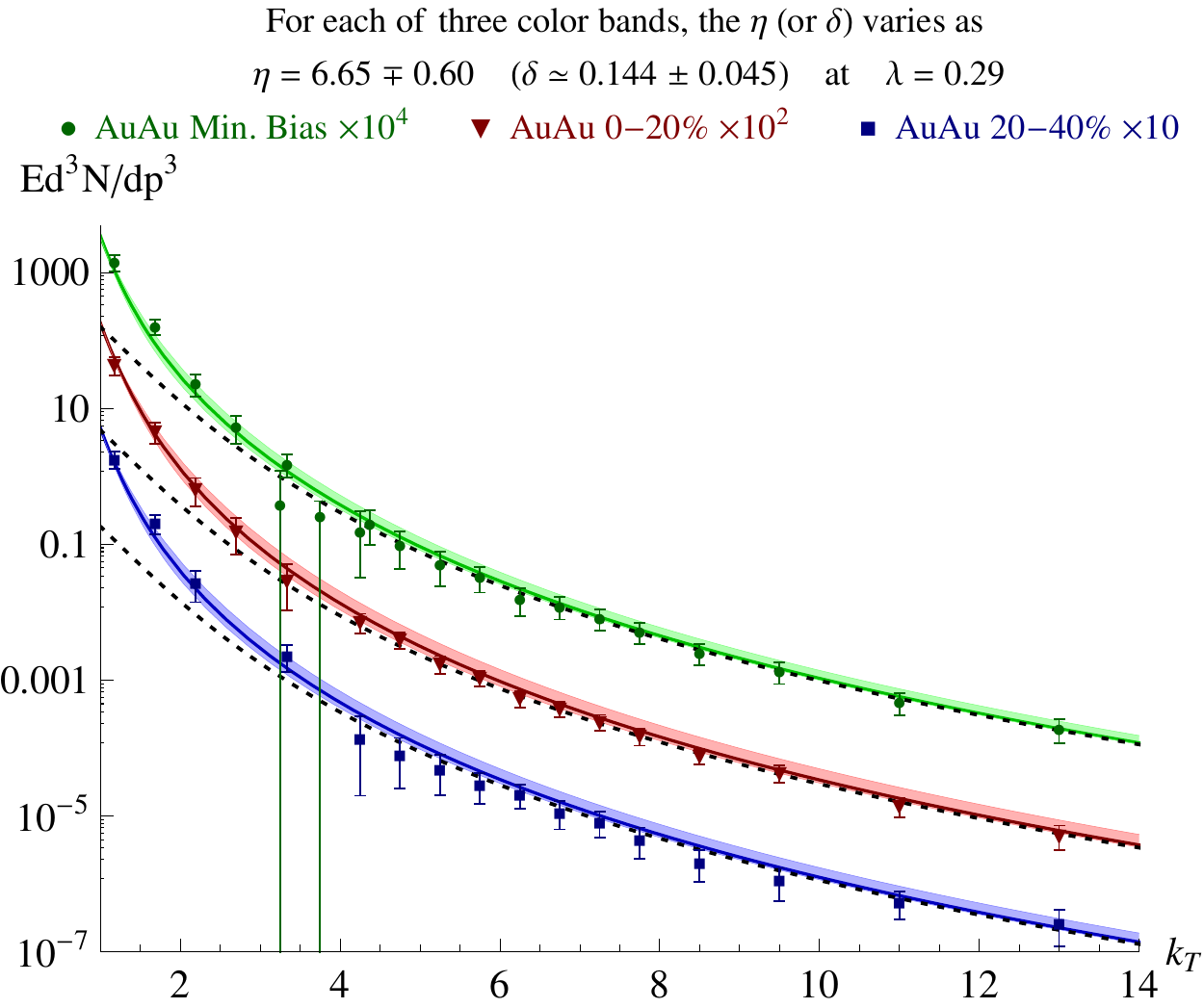} }
\end{center}
\caption{Comparison between the PHENIX photon data and the present model with $\lambda = 0.29$ and varied values of $\eta$ for three centrality bins.  The black dashed curves represent the $T_{AA}$-scaled pp yield from Eq.(\ref{eqn_G}), and the colorful bands represent full yield including also the Glasma contribution from Eq.(\ref{eqn_F}) with the upper and lower boundary curves for each band corresponding to the results with the parameter $\eta=6.05$ and $\eta=7.25$, respectively  (see text for more details). }
\label{photons}
\vspace{0.2cm}
\end{figure}

Finally we present our fitting results for the PHENIX photon data. In  Fig.\ref{photons_chi}, we show the $\chi^2$/d.o.f. analysis in the $\lambda-\eta$ parameter space for the PHENIX photon data in three centrality bins by plotting the three contours corresponding to $1$-$\sigma$(blue), $2$-$\sigma$(green), and $3$-$\sigma$(red) errors. Based on this analysis, one can identify
 a best-fitting zone (at about $2-\sigma$ level) to be $\lambda=0.29\pm 0.05$ and $\eta=6.65\mp 0.60$ --- the latter corresponding to the asymmetry parameter $\delta=0.144\pm 0.045$.

Having identified the optimal parameter regime, we now show in Fig.\ref{photons} the direct comparison between data and our model fitting with $\lambda=0.29$ and $\eta=6.65\mp 0.60$. For each centrality bin, the PHENIX data points are compared with: (a) the contribution from only the $T_{AA}$-scaled pp yield, i.e. the Hagedorn function in Eq.(\ref{eqn_G}), represented by the black dashed curves; and (b) the full yield including both the $T_{AA}$-scaled pp yield and the contribution from the Glasma in Eq.(\ref{eqn_F}), represented by the colorful bands where the upper and lower boundary curves for each band correspond to the results with $\eta=6.65-0.60=6.05$ and $\eta=6.65+0.60=7.25$, respectively. The plots show good agreement between the PHENIX data and our model fitting at all centralities.

The only parameter directly determined from fitting is the overall normalization $C_\gamma$ in Eq.(\ref{eqn_F}): it has been fixed from
the $0$-$20^{\circ\!\!}/_{\!\!\circ}$ case to be $C_\gamma \approx 0.0234 fm^2$   and then used in all other centralities. We notice that this value is consistent with the expectation $C_\gamma \sim \alpha R_0^2$ provided $\alpha=1/137$, $R_0$ of the order of a  $\rm fm$ and a reasonable coefficient.

A few remarks are in order from the comparison. First, while the very high $k_T>3\, \rm GeV$ data are well described by the $T_{AA}$-scaled pp yield only, the inclusion of the Glasma contribution is necessary and even dominant for describing the ``excess'' in yield and the $k_T$-dependence in the softer region about $1\sim 3\, \rm GeV$. Second, the fact that data for varied centralities can be well fit by one parameter $C_\gamma$ fixed at one centrality provides strong evidence that our model for Glasma photon production has captured the essential geometrical scaling in such data in the relatively lower-$k_T$ region. Last, the comparison implies for the parameter $\eta$ a preferred region $\eta=6.65\mp 0.60$, corresponding to a region $\delta=0.144\pm 0.045$ for the asymmetric parameter $\delta$ in Eq.(\ref{eqn_PL}) which appears to indicate strong anisotropy between longitudinal and transverse scales in the Glasma evolution.

An issue we cannot address at present is the fact that the photon excess measured at PHENIX has reasonably large $v2$ \cite{Kistenev:2011zz} since in our computations we cannot account for the effects of transverse expansion.  Although we expect sizable flow to be developed in the Glasma, it remains an issue to be seen if it is sufficient to explains the data, or if there is new physics involved\cite{Basar:2012bp}.

\subsection{ Dileptons}

Consider the dilepton differential yield in both invariant mass $m_{ee} \equiv M$ and transverse momentum $k_T$.  The phenomenological formula is of the form
\begin{equation}
{dN_{C \rightarrow DY} \over d^{2}k_{T}dy\,dM^{2}} = C_{ll}
N_{part}^{2/3}\Lb {Q_{sat} \over M} \Rb^{\eta^{\prime}}{e^{-k_{T}/\mu} \over \mu^{2}}\,,
\label{eqn_dilep1}
\end{equation}
As before, we use the following expression for the saturation momentum:
\begin{equation}
Q_{sat}^{2}(k_{T}/\sqrt{s}) =  Q_{0}^{2}\Lb \sqrt{s}\times 10^{-3} \over {k_{T}}
\Rb^{\lambda}e^{\lambda y}\,.
\label{eqn_mom3}
\end{equation}
which includes its evolution with the transverse momentum $k_{T}$ as well as the rapidity $y$. 

The data we aim to describe are for $k_{T}$-dependence of  $e^{+}e^{-}$ pairs at various given mass bins, as measured by PHENIX in $Au+Au$ \,$200\,GeV$ minimum bias collisions.  For a given mass bin $[M_{min}\,,M_{max}]$, we can integrate the differential yield to obtain the $k_T$-spectra:
\begin{equation}
{dN_{C \rightarrow DY} \over 2\pi k_{T}\,dk_{T}dy} =  C_{ll}
N_{part}^{2/3}\,Q_{sat}^{\eta^{\prime}}{e^{-k_{T}/\mu} \over \mu^{2}}
\int_{M_{min}}^{M_{max}}{2M \over M^{\eta^{\prime}}}dM\,,
\end{equation}
which eventually leads to
\begin{eqnarray}
{1 \over (N_{part}/2)}{dN_{C \rightarrow DY} \over 2\pi k_{T}\,dk_{T}dy} & = & {4 C_{ll}  \over (2 - \eta^{\prime})}\,N_{part}^{-1/3} \times
\Lb M_{max}^{2 - \eta^{\prime}} - M_{min}^{2 - \eta^{\prime}} \Rb
\nonumber\\
& \times &  \frac{e^{-k_{T}/\mu}}{\mu^{2}} \times \Lb Q_{0}^{2}\Lb \sqrt{s}\times 10^{-3} \over {k_{T}}
\Rb^{\lambda} \Rb^{\eta^{\prime}/2}\,.
\label{eqn_dilep2}
\end{eqnarray}
In the above we have also incorporated Eq.(\ref{eqn_mom3}).
At this point it shall be emphasized that in our model the dileptons generated from the Glasma shall have their mass bounded by the in-medium mass of gluons in the Glasma, and therefore our formula shall not be applied to too small values for the pair mass. Accordingly, we focus on comparison with data in a mass regime $0.2 ~ GeV \le M \le 1~GeV$. Since the parameters $\lambda$ and $\eta'$ (determined by $\delta$) are well constrained from the photon fitting, we will use those optimal values implied by the photon data also for the dilepton fitting, i.e. $\lambda=0.29$ and $\eta'=4.73\pm 0.20$ (corresponding to $\delta=0.144\mp 0.045$).

\begin{figure}[h!]
\vspace{0.0cm}
\begin{center}
\scalebox{1.0}[1.0] {
\hspace{0.2cm}
  \includegraphics[scale=.80]{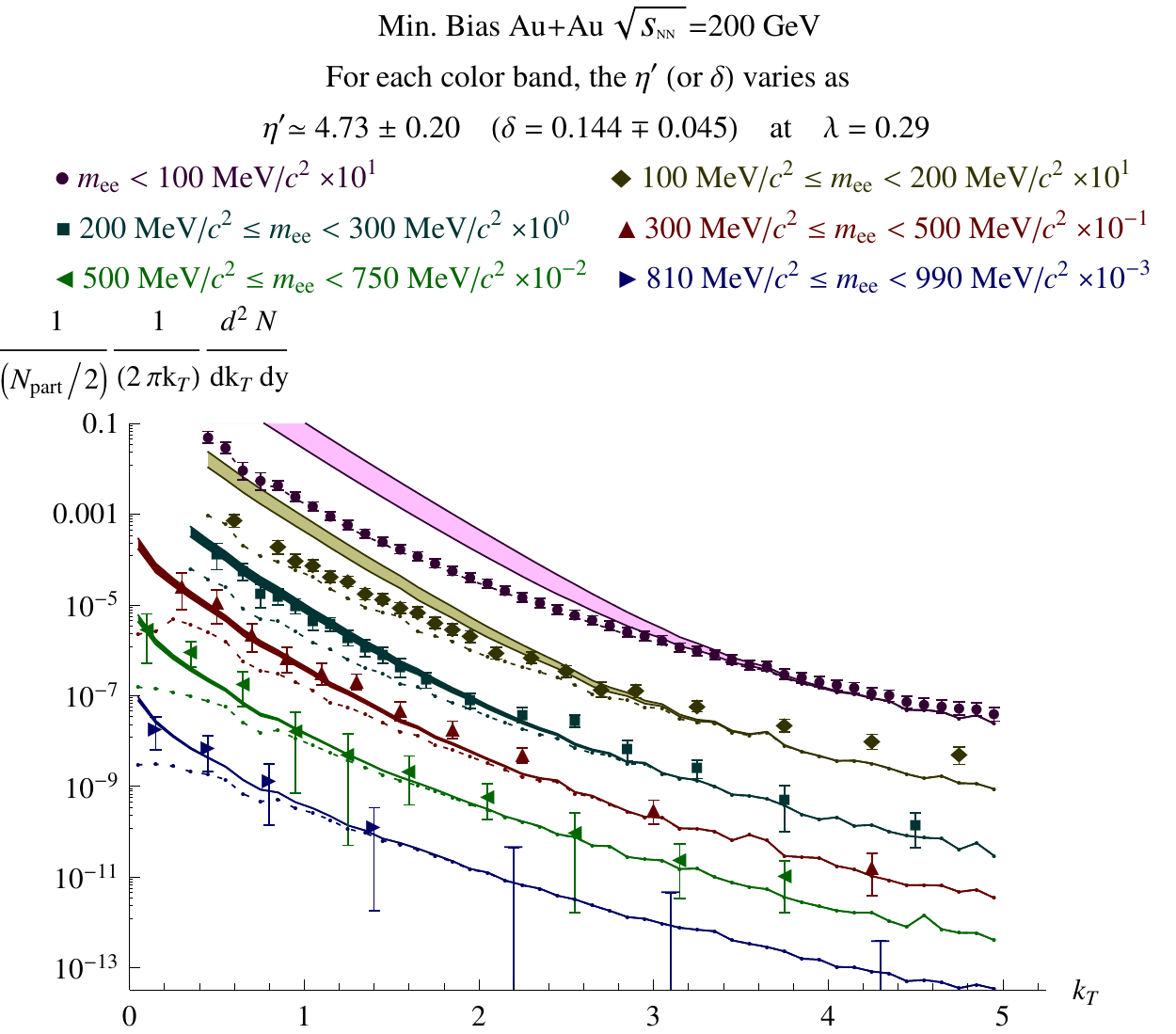} }
\end{center}
\caption{(Color online) The comparison between the PHENIX dilepton data and our present model with parameters $\lambda = 0.29$ and $\eta'=4.73\pm0.20$ for the four higher mass bins.  The dashed curves represent the background contributions from the hadronic cocktail and charmed mesons, and the colorful bands represent the full yield including also the Glasma contribution from Eq.(\ref{eqn_dilep2}) with the upper and lower boundary curves for each narrow band corresponding to the results with the parameter $\eta'=4.93$ and $\eta'=4.53$, respectively  (see the text for more details).}
\label{dileptons}
\vspace{0.2cm}
\end{figure}

For each mass bin of the dilepton yield, the dashed curves represent the background yield formed by contributions of the hadronic decay cocktail and charmed mesons. For the four higher mass bins relevant to the production from the Glamsa, we add on top of the background the additional contribution given by Eq.\,(\ref{eqn_dilep2}): the results for the total yield are represented by the colorful bands where the upper and lower boundary curves for each band correspond to the results with the parameter $\eta'=4.93$ and $\eta'=4.53$, respectively. For the $k_T$ width we have found the optimal value $\mu=0.2\,GeV$. The fitting quality has a relatively strong dependence on $\mu$. The optimal value for the overall normalization $C_{ll}$ determined from such fitting is $C_{ll}\approx 3.5 \times 10^{-6}\,GeV^{-2}$. This coefficient $C_{ll}$ is plausibly expected to be parametrically much smaller than the $C_{\gamma}$ in the photon case, as the former has one more power in its dependence on the electromagnetic coupling $\alpha=1/137$. Our fitting result for the PHENIX dilepton data is shown in Fig.\,\ref{dileptons}.

Let us further examine the dependence of dilepton yield on the pair invariant mass $M$ in the Glasma scenario. To do that, we shall first model the $e^{+}e^{-}$ pair production from various hadron decays using the fast Monte-Carlo decay generator called EXODUS. It is a phenomenological event generator that allows to simulate the phase-space distribution of all relevant sources of electrons, electron pairs and the decay of these sources. Also, it allows one to include the filtering for the detector geometrical acceptance and the detector resolution. The relevant primary hadrons (mesons) that involve electrons in the final state are $\pi^{0}$, $\eta$, $\eta^{\prime}$, $\rho$, $\omega$, $\phi$, $J/\Psi$ and $\Psi^{\prime}$. It is assumed that all the mesons have a constant rapidity density in the range $|\Delta y| \leq 0.35$, and a uniform distribution in azimuthal angle. By this way we obtain the summed up hadronic cocktail consisted of dileptons from all these mesons. The next step is to add the charm contribution to the cocktail (the same procedure was applied for computations shown in Fig.\,\ref{dileptons} as well). Then we treat the Glasma as a source of pair generation, which can be incorporated into EXODUS using Eq.\,(\ref{eqn_dilep1}) as a general formula for the pair production. Starting from  Eq.\,(\ref{eqn_dilep1}) one can get separate formulae for dilepton yield as functions of $M$, $y$ and $k_{T}$, respectively,  by integrating Eq.\,(\ref{eqn_dilep1}) over specific kinematic intervals, i.e. 
\begin{itemize}
\item[a)] for ${k_{T,min}} \le k_{T} \le {k_{T,max}}$ and $-Y_0 \le y \le Y_0$.
\item[b)] for ${k_{T,min}} \le k_{T} \le {k_{T,max}}$ and ${M_{min}} \le k_{T} \le {M_{max}}$.
\item[c)] for ${M_{min}} \le k_{T} \le {M_{max}}$ and $-Y_0 \le y \le Y_0$.
\end{itemize}
These formulae are shown as follows:
\begin{eqnarray}
\!\!\!\!\!\!\!{dN_{C \rightarrow DY} \over dM} & = & \!\!\! 4\pi\, C_{ll} \,
N_{part}^{2/3}\Lb Q_{0}^{2}\Lb \sqrt{s}\times 10^{-3} \Rb^{\lambda} \Rb^{\eta^{\prime}/2}\,\times
\nonumber\\
& \times & \!\!\! {\Lb 1 \over \mu \Rb}^{(\lambda\eta^{\prime}/2)}\!\!\left[ \Gamma\!\Lb 2 -
0.5\lambda\eta^{\prime},{k_{T,min} \over \mu} \Rb - \Gamma\!\Lb 2 - 0.5\lambda\eta^{\prime},{k_{T,max}\over \mu}
\Rb \right]\,\times
\nonumber \\
& \times & \!\!\!  {4\sinh{\![(\lambda\eta^{\prime}/2)Y_{0}]} \over \lambda\eta^{\prime}} \Lb {1 \over M^{\eta^{\prime} - 1}} \Rb\,,
\label{eqn_dilep4}
\end{eqnarray}
\begin{eqnarray}
\!\!\!\!\!\!\!{dN_{C \rightarrow DY} \over dy} & = & \!\!\! 2\pi\, C_{ll} \,
N_{part}^{2/3}\Lb Q_{0}^{2}\Lb \sqrt{s}\times 10^{-3} \Rb^{\lambda} \Rb^{\eta^{\prime}/2}\,\times
\nonumber\\
& \times & \!\!\! {\Lb 1 \over \mu \Rb}^{(\lambda\eta^{\prime}/2)}\!\!\left[ \Gamma\!\Lb 2 -
0.5\lambda\eta^{\prime},{k_{T,min} \over \mu} \Rb - \Gamma\!\Lb 2 - 0.5\lambda\eta^{\prime},{k_{T,max}\over \mu}
\Rb \right]\,\times
\nonumber \\
& \times & \!\!\! \Lb {2 \over\eta^{\prime} -2} \Rb \Lb M_{min}^{2 - \eta^{\prime}} - M_{max}^{2 - \eta^{\prime}} \Rb\,
e^{-{\lambda\eta^{\prime} \over 2}|y|}\,,
\label{eqn_dilep5}
\end{eqnarray}
\begin{eqnarray}
\!\!\!\!\!\!\!{dN_{C \rightarrow DY} \over dk_{T}} & = & \!\!\! 2\pi\, C_{ll} \,
N_{part}^{2/3}\Lb Q_{0}^{2}\Lb \sqrt{s}\times 10^{-3} \Rb^{\lambda} \Rb^{\eta^{\prime}/2}\,\times
\nonumber\\
& \times & \!\!\! {4\sinh{\![(\lambda\eta^{\prime}/2)Y_{0}]} \over \lambda\eta^{\prime}} \Lb {2 \over\eta^{\prime} -2} \Rb \Lb M_{min}^{2 - \eta^{\prime}} - M_{max}^{2 - \eta^{\prime}} \Rb \Lb k_{T}^{1 - {\lambda\eta^{\prime} \over 2}}
{e^{-{k_{T} \over \mu}} \over \mu^{2}} \Rb\,.
\label{eqn_dilep6}
\end{eqnarray}
Afterwards we incorporate these formulae (describing the contribution from the Glasma) into EXODUS, which in particular allows us to obtain the mass spectrum of the $e^{+}e^{-}$ pairs filtered through the acceptance of the PHENIX detector. 

\begin{figure}[h!]
\vspace{0.0cm}
\begin{center}
\scalebox{1.0}[1.2] {
\hspace{-0.2cm}
  \includegraphics[scale=0.75]{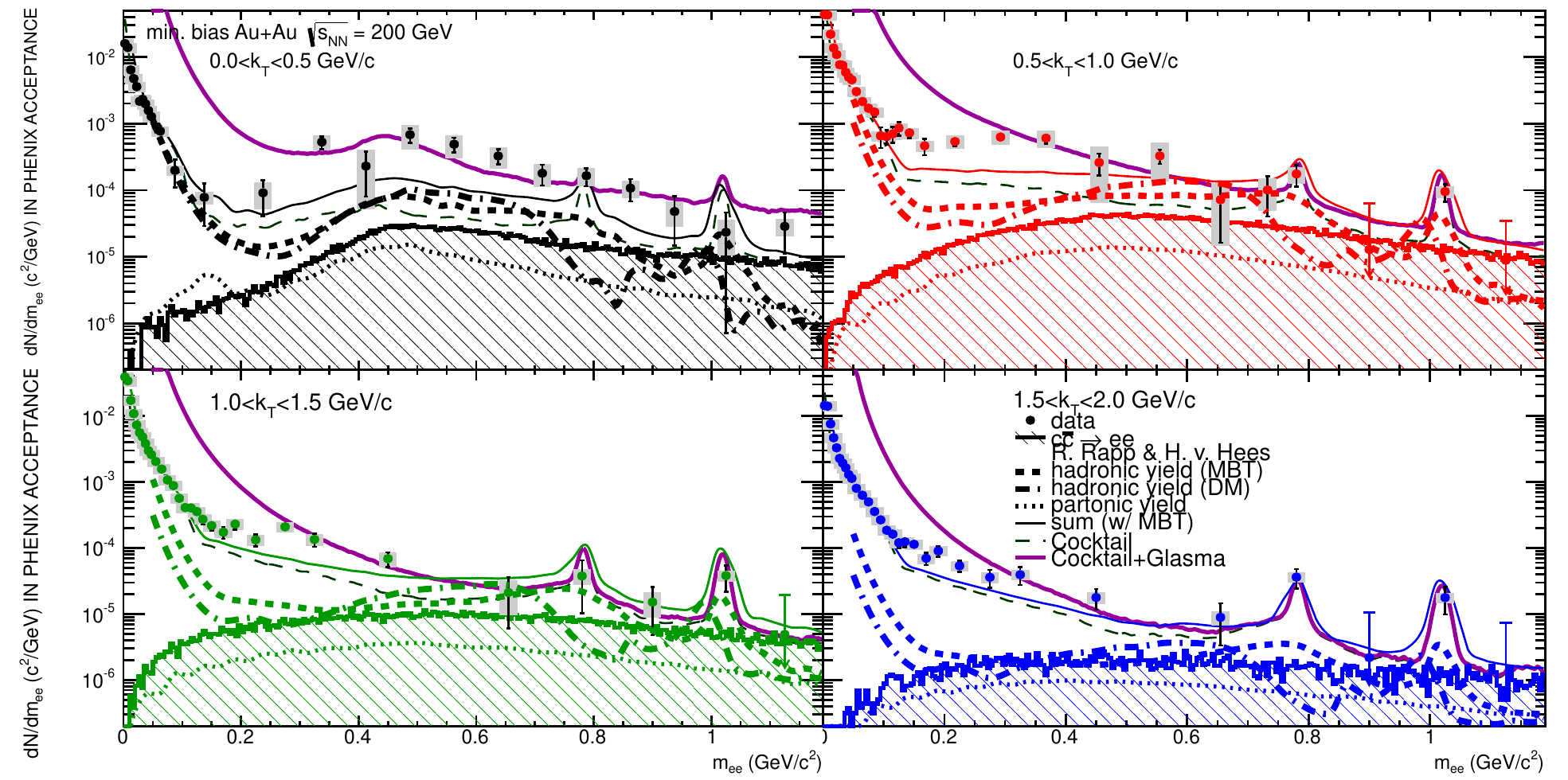} }
\end{center}
\caption{(Color online) Invariant mass spectra of the $e^{+}e^{-}$ pairs in Min. Bias $Au + Au$ collisions for different $p_{T}$ windows compared to the expectations from the calculations of Rapp and van Hees (\cite{Rapp1}, \cite{rapp_RHIC}, \cite{vHvRapp}), separately showing the partonic and hadronic yields in different scenarios for the $\rho$ spectral function, namely the so called ``Hadron Many Body Theory'' (HMBT) and ``Dropping Mass" (DM). The calculations have been added to the cocktail of hadronic decays (where the contribution of the freeze-out $\rho$ meson is subtracted) and charmed meson decay products.  This figure is from Ref.\,\cite{Adare:2009qk} where we have also added the hadronic cocktail and charmed meson contributions (called Cocktail) plus that from the Glasma. In our Cocktail the $\rho$ meson contribution is included.}
\label{Acceptance_Dileptons_mee1}
\vspace{0.2cm}
\end{figure}

\begin{figure}[h!]
\vspace{0.0cm}
\begin{center}
\scalebox{1.0}[1.2] {
\hspace{-0.2cm}
  \includegraphics[scale=0.75]{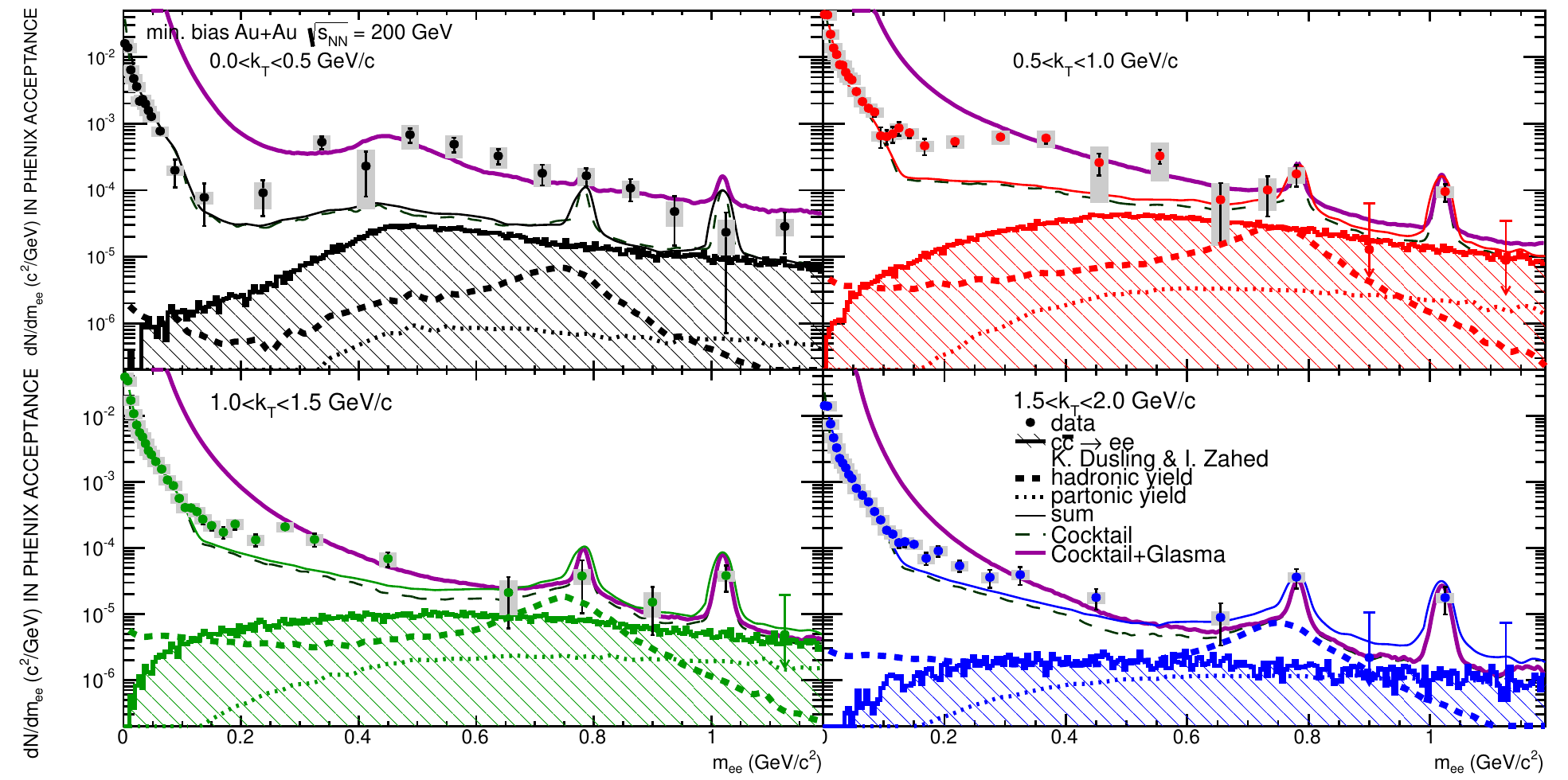} }
\end{center}
\caption{(Color online) Invariant mass spectra of the $e^{+}e^{-}$ pairs in Min. Bias $Au + Au$ collisions for different $p_{T}$ windows compared to the expectations from the calculations of Dusling and Zahed (\cite{dusling_RHIC}, \cite{dusling_mod}), separately showing the partonic and the hadronic yields. The calculations have been added to the cocktail of hadronic decays (where the contribution of the freeze-out $\rho$ meson is subtracted) and charmed meson decays products.  This figure is from Ref.\,\cite{Adare:2009qk} where we have also added the hadronic cocktail and charmed meson contributions (called Cocktail) plus that from the Glasma. In our Cocktail the $\rho$ meson contribution is included.}
\label{Acceptance_Dileptons_mee2}
\vspace{0.2cm}
\end{figure}

\begin{figure}[h!]
\vspace{0.0cm}
\begin{center}
\scalebox{1.0}[1.2] {
\hspace{-0.2cm}
  \includegraphics[scale=0.75]{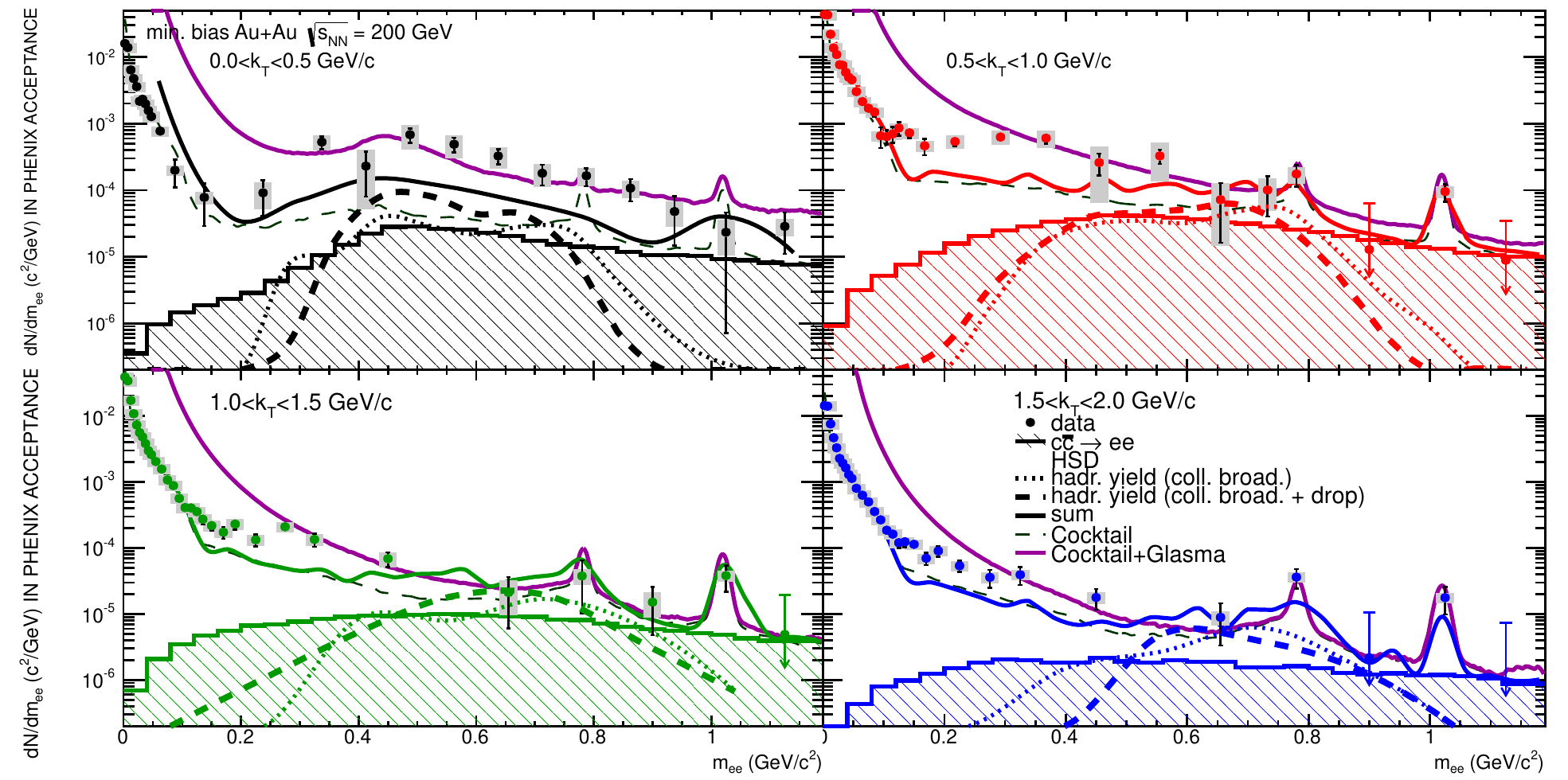} }
\end{center}
\caption{(Color online) Invariant mass spectra of the $e^{+}e^{-}$ pairs in Min. Bias $Au + Au$ collisions for different $p_{T}$ windows compared to the expectations from the calculations of Cassing and Bratkovskaya (\cite{cassing_RHIC}, \cite{cassing0},\cite{cassing_HSD}), separately showing the partonic and hadronic yields calculated with different implementations of the $\rho$ spectral function, namely according to collisional broadening, with or without a dropping mass scenario. The calculations which include the dropping mass scenario have been added to the cocktail of hadronic decays (which is calculated by the HSD model itself) and charmed meson decays products. This figure is from Ref.\,\cite{Adare:2009qk} where we have also added the hadronic cocktail and charmed meson contributions (called Cocktail) plus that from the Glasma. In our Cocktail the $\rho$ meson contribution is included.}
\label{Acceptance_Dileptons_mee3}
\vspace{0.2cm}
\end{figure}

In Figs.\,\ref{Acceptance_Dileptons_mee1}, \ref{Acceptance_Dileptons_mee2} and \ref{Acceptance_Dileptons_mee3} we show the $e^{+}e^{-}$ invariant mass spectra in different $k_{T}$ windows from data compared to the sum of the background (the cocktail including the charm) and the Glasma. Also shown is the comparison with the dilepton yields obtained by Rapp and van Hees (\cite{Rapp1}, \cite{rapp_RHIC}, \cite{vHvRapp}) in Fig. ,\ref{Acceptance_Dileptons_mee1}, Dusling and Zahed (\cite{dusling_RHIC}, \cite{dusling_mod}) in Fig. \ref{Acceptance_Dileptons_mee2}. , and Cassing and Bratkovskaya (\cite{cassing_RHIC}, \cite{cassing0},  \cite{cassing_HSD}) in Fig.  \ref{Acceptance_Dileptons_mee3}, respectively. In their simulations the contribution from the hadronic and partonic medium, and the charm expectations are shown separately. In these papers, in-medium modifications of the $\rho$ meson spectral function are used, which could be responsible for the enhancement of the dilepton yield below the $\rho$ mass (for more details see the page $42$ of Ref.\,\cite{Adare:2009qk}). Note that the cocktail plus the Glasma dilepton yield in Figs.\,\ref{Acceptance_Dileptons_mee1}, \ref{Acceptance_Dileptons_mee2} and \ref{Acceptance_Dileptons_mee3} is consistent with that from  Fig.\,\ref{dileptons}  within the mass range of \,$0.3~GeV \le M \le 1~GeV$. The general observation from all these comparisons with data is that in the aforementioned mass range the estimates from Glasma plus cocktail contributions agree with data  reasonably well and also nicely describe the $k_T$ spectrum reproducing the dominant excess from low $k_T$ region.

\section{Conclusion and Discussion}

The conclusion of this paper is that  the photons and dileptons data measured at PHENIX may be consistent with a scenario that assumes important contribution to the electromagnetic production during the pre-equilibrium stage based on the Glasma hypothesis with the existence of a condensate.  There are of course alternative explanations.  We briefly outline below some of the places where alternative hypothesis may be viable or attractive and where further investigations would be desired.

First, we discuss the results for the photon production.  The features that can be explained naturally within the Glasma scenario include the shape of the $k_T$ spectrum as well as its centrality dependence. 
However:
\begin{itemize}

\item{The photon $k_T$ spectrum can be explained by hydrodynamic expansion of a very hot quark-gluon plasma. The problem with that  explanation is that it requires thermalization at quite early times (such that the system is still hot), while there is now accumulating evidences  that  the  system at so early a stage is more of a Glasma.}

\item{The photon spectrum satisfies geometric scaling across centrality.  This arises naturally from the Glasma approach.  Nevertheless
such scaling properties are characteristic of hadronic processes, and can be built into hydrodynamic computations by choice
of Glasma-like initial conditions.}

\item{The photon spectrum has large elliptic flow (i.e. significant azimuthal anisotropy).  This feature poses serious and generic difficulty for  both a thermalized QGP and a Glasma description  if one tries to generate the flow during the early time expansion.  The problem might be a little less severe for the Glasma as the quarks become substantial in Glasma only toward the later stage in the lifetime of the Glasma.   One may presumably expect that either there are yet-to-be-understood generation of pre-equilibrium flow at early times, or the source of the photons might be from some entirely different source.}

\end{itemize}

Let us then turn to the dilepton production. By including the Glasma contribution on top of the conventional ``cocktail'' production,  a good description of the mass and $k_T$ dependence of the dilepton spectrum has been achieved. In particular the excess dominantly at small $k_T$ may be linked with production from the Bose condensate in the Glasma.  However:
\begin{itemize}

\item{So far a consensus is lacking between the PHENIX and STAR experiments over the magnitude of the dilepton enhancement as well as on the dominant $k_T$ range of such enhancement. In addition, the largest barrier to a description of this enhancement in terms of the thermal QGP production is that it cannot reproduce the magnitude of the effect.  In the present Glasma estimates,  the magnitude of the effect is not computed.}

\item{Even if it could be experimentally settled that the source of the dilepton excess arises from low $k_T$, it could be that such low $k_T$ dileptons may arise via a similar mechanism from some other sort
of condensate not associated with the Glasma.  In addition, the existence and properties of a transient Bose condensate in the Glasma  are still under intensive scrutiny. }

\item{The Glasma hypothesis predicts geometric scaling of the dilpton spectrum, but a critical verification of such predictions will have to wait for  sufficient data in the future.}\\
\end{itemize}

Bearing all these caveats in mind, we emphasize again the valuable lessons from this study as we have already listed in the Introduction. It should also be emphasized that the presented study of the Glasma scenario here is not intended to draw any conclusion, on the basis of a comparison with PHENIX data, regarding whether the Glasma or the thermalized Quark-Gluon Plasma scenario is favored. Instead, the primary point is a theoretically well motivated expectation, i.e. for a considerable window in the early time evolution of the system, the matter produced in heavy ion collisions will be better described by a thermalizing Glasma rather than a thermalized Quark-Gluon Plasma. Based on that expectation we have set out in this paper to examine its implications for the photon and dilepton production, and have found a plausible  phenomenological description of data. Admittedly like all such first attempts, there is much uncertainty.  One however may remember the significant time and efforts it took to refine the thermal photon and dilepton computations and to achieve some consensus over theoretical predictions. In reflection of this history, we hope that the first small steps taken here will lead toward a deeper understanding of what we consider to be an important question:\\
{\em \indent  When and for what phenomena is the strongly interacting Quark-Gluon Plasma better described as a thermalized Quark-Gluon Plasma or as a thermalizing Glasma?}

\section*{Acknowledgements}
%%%%%%%%%%%%%%%%%%%%%%%%%%
The research of M. Chiu is supported under DOE Contract No. DE-AC02-98CH10886. The research of T. K. Hemmick and V. Khachatryan is supported under DOE Contract No. DE-FG02-96ER40988. A. Leonidov acknowledges support from the RFBR grant 12-02-91504-CERN and RAS LHC program. The research of J. Liao and L. McLerran is supported under DOE Contract No. DE-AC02-98CH10886. L. McLerran thanks the Theoretical Physics Institute of the  University of Heidelberg where this work was in part developed.  He is supported there as a Hans Jensen Professor of Theoretical Physics. J. Liao is grateful to RIKEN BNL Research Center for partial support.

\section*{Appendix}

\appendix

\section{Photon emission from glasma}\label{apphot}

The standard expression for the fixed-box photon production rate from the Compton channel $gq \to \gamma q$ reads:
\begin{equation}
E\dfrac{dN}{d^4x d^3p} \propto F_q (E/\Lambda) \frac{1}{E} \int_{\mu^2}^\infty ds \; (s - \mu^2) \; \sigma_{gq \to \gamma q}(s) \;
\int_{s/4E}^\infty dE_g f_g (E_g) \left[ 1 - F_q (E_g/\Lambda) \right] \label{photfixbox}
\end{equation}
where $F_q (E/\Lambda)$ is the quark distribution function, $f_g$ is the gluon distribution function, the lower limit for integration over gluon
energy $E_g$ follows from kinematics, $\mu^2$ is an infrared cutoff needed to regularize the $t(u)$ - channel singularity for diagrams with
massless particle exchange which in our case is the Debye mass  $\mu^2=\Lambda \Lambda_s$ and $\sigma_{gq \to \gamma q} (s)$ is the cross-section for gluon Compton effect $gq \to \gamma q$.

In the high energy limit and for small quark densities $F_q$ Eq. (\ref{photfixbox}) simplifies to
\begin{equation}
E\dfrac{dN}{d^4x d^3p} \propto F_q (E/\Lambda) \frac{\Lambda_s \Lambda}{E} \int_{1}^\infty dy \; \ln y \;
\int_{\frac{y \Lambda_s \Lambda}{4E}}^\infty dE_g f_g (E_g)  \label{photfixbox1}
\end{equation}
To elucidate the scaling dependencies of (\ref{photfixbox1}) let us consider the following simple parametrization of the gluon density $f_g$:
\begin{eqnarray}
f_g (E_g) & = & {\rm const.}, \;\;\;\;\;\;\;\;\;\;\; E_g<\Lambda_s \nonumber \\
f_g (E_g) & = & {\rm const.} \frac{\Lambda_s}{E_g}, \;\;\;\;\;\; \Lambda_s<E_g<\Lambda \nonumber \\
f_g (E_g) & = & 0 , \;\;\;\;\;\;\;\;\;\;\;\;\;\;\;\;\;\;\; E_g>\Lambda
%f_g (E_g) & = & \frac{\Lambda_s}{\Lambda} \exp (-E_g/\Lambda), \;\;\;\;\;\; E_g>\Lambda
\end{eqnarray}
Let
\begin{equation}
I(y, \vert \Lambda_s,\Lambda,E) = \int_{\frac{y \Lambda_s \Lambda}{4E}}^\infty dE_g f_g (E_g) \equiv \int_{E^*}^\infty dE_g f_g (E_g)
\end{equation}
Then
\begin{equation}
I= \Lambda_s \left[ \; \theta(\Lambda_s-E^*) I_1 + \theta(\Lambda-E^*) \theta(E^*-\Lambda_s) I_2 \; \right]
\end{equation}
where
\begin{eqnarray}
I_1 & = & 1-y \frac{\Lambda}{4 E} + \ln \left(\frac{\Lambda}{\Lambda_s} \right)  \nonumber \\
I_2 & = & \ln \left( \frac{\Lambda}{\Lambda_s} \right) + \ln \left(\frac{4 E}{\Lambda} \right) - \ln y
\end{eqnarray}
and, therefore, 
\begin{equation}
E\dfrac{dN}{d^4x d^3p} \propto F_q (E/\Lambda) \frac{\Lambda_s^2 \Lambda}{E} 
\left[ \int_1^{\frac{4E}{\Lambda}} dy \ln y I_1 + \int_{\frac{4E}{\Lambda}}^{\frac{4E}{\Lambda_s}} dy \ln y I_2 \right] \label{photfixbox3}
\end{equation}
A straightforward calculation shows that in the limit of $\Lambda \gg \Lambda_s$ the leading contribution is coming from the second term in (\ref{photfixbox3}) and is proportional to $(E/\Lambda_s) \phi(E/\Lambda)$ up to log accuracy, where $\phi(E/\Lambda)$ is some easily analytically calculable function, so that the final result for the rate reads:
\begin{equation}
E\dfrac{dN}{d^4x d^3p} \propto F_q (E/\Lambda) \Lambda \Lambda_s \phi(E/\Lambda), \label{photfixbox4}
\end{equation}
The expression (\ref{photfixbox4}) is exactly of the form conjectured in (\ref{photratefb}).

\section{Dilepton emission from glasma}\label{apdil}

The expression for the static rate of production of dilepton pairs with invariant mass $M$ for massless quarks and leptons reads
\begin{eqnarray}
\dfrac{dN^{l^+l^-}}{d^4 x d M^2} & \sim & M^2 \sigma_{q{\hat q} \to l^+l^-} (M^2) \int_0^\infty dE_q F_q (E_q/\Lambda) \int_{M^2/4E_q}^\infty dE_{\bar q} F_{\bar q} (E_{\bar q}/\Lambda)
\nonumber \\
& = & M^2 \sigma_{q{\hat q} \to l^+l^-} (M^2)\, \Lambda^2 \int_0^\infty dy F_q(y) \int_{M^2/4 \Lambda^2 y} dx F_q(x) \label{dil1}
\end{eqnarray}
Taking into account that $M^2 \sigma_{q{\hat q} \to l^+l^-} (M^2) \sim \;{\rm const.}$, we can already see the scaling behavior of the static dilepton production rate (\ref{dilratefb}), i.e. 
\begin{eqnarray}
\dfrac{dN^{l^+l^-}}{d^4 x d M^2} & \sim & \Lambda^2 \, \Phi(M/\Lambda) 
\end{eqnarray}
with $\Phi(M/\Lambda)\equiv  \int_0^\infty dy F_q(y) \int_{(M/\Lambda)^2/4y} dx F_q(x)$. 
To further explicitly demonstrate  the scaling behavior of the static dilepton production rate (\ref{dilratefb}), let us consider two examples with explicit forms of quark distribution function. 

First let us consider a simple hard-cutoff quark distribution function $F_q=\theta(\Lambda - E)$. In this case we can easily obtain 
\begin{eqnarray}
\dfrac{dN^{l^+l^-}}{d^4 x d M^2} & \sim & \Lambda^2 \left[ 1-\frac{(M/\Lambda)^2}{4} + \frac{(M/\Lambda)^2}{4} \ln \frac{(M/\Lambda)^2}{4}  \right]
\end{eqnarray}

Second let us consider an exponential quark distribution function $F_q=\exp (-E/\Lambda)$. In this case  we get the following result
\begin{eqnarray}
\dfrac{dN^{l^+l^-}}{d^4 x d M^2} & \sim & \Lambda^2 \, \sqrt{\dfrac{M^2}{\Lambda^2}}  K_1 \left( \dfrac{M}{\Lambda} \right) \nonumber \\
& = & M \Lambda K_1 \left( \dfrac{M}{\Lambda} \right) \equiv \Lambda^2 \, \left[ \dfrac{M}{\Lambda} K_1 \left( \dfrac{M}{\Lambda} \right) \right] \label{dil2}
\end{eqnarray}
where $K_1 \left( \dfrac{M}{\Lambda} \right)$ is a Bessel function.  The expression for the rate (\ref{dil2}) is exactly of the form conjectured in (\ref{dilratefb}). Although the derivation used a particular form of quark distribution, it is easy to prove that this result is generic. For $\Lambda = T$, (\ref{dil2}) is exactly the standard fixed box thermal rate.

\end{document}